%% file: ACC_main.tex
\documentclass[twocolumn,10pt]{IEEEtran}
\usepackage{ifpdf, flushend,subfigure}
\usepackage{graphicx}
\usepackage{epstopdf}
\graphicspath{{../}}
\usepackage{multirow}
\usepackage{float}
\usepackage[cmex10]{amsmath}
\usepackage {amssymb}
\usepackage{array}
\usepackage{mdwmath}
\usepackage{mdwtab}
\usepackage{eqparbox}
\usepackage{url}
\usepackage{nomencl}
\usepackage{cite}
\usepackage{algorithm}
\usepackage{algorithmic}
\usepackage{makeidx}
\usepackage{ifthen}
\usepackage{subfigure}
\usepackage{caption}
\usepackage{pdflscape}
\usepackage{hyperref}
\usepackage{xcolor}

\def\mbi{\mathbb{I}}

\def\ella#1{{\color{black}#1}}
 
\def\an#1{{\color{black}#1}}

\makeatletter
\def\munderbar#1{\underline{\sbox\tw@{$#1$}\dp\tw@\z@\box\tw@}}
\makeatother

\input{slide_macros.tex}

\definecolor{mygreen}{rgb}{0.0, 0.5, 0.0}
\newcommand{\mattiasay}[1]
{\color{black}{#1}\color{black}}

\begin{document}

\title{
%Distributed Nash Equilibrium Seeking over Directed Graphs with Row-Stochastic Matrices
%Nash equilibrium seeking over digraphs with network-independent step-sizes
Nash equilibrium seeking over digraphs with row-stochastic matrices and network-independent step-sizes
% \mattiasay{Nash equilibrium seeking over row-stochastic digraphs with network independent step-sizes}
}

\author{Duong Thuy Anh Nguyen\textsuperscript{1}, Mattia Bianchi\textsuperscript{2}, Florian Dörfler\textsuperscript{2},
Duong Tung Nguyen\textsuperscript{1}, Angelia Nedi\'c\textsuperscript{1}
\thanks{1. School of Electrical, Computer and Energy Engineering, Arizona State University, Tempe, AZ, United States. 
Email: \{dtnguy52,~duongnt,~Angelia.Nedich\}@asu.edu. This  work was supported by the NSF award CCF-2106336. 
2. Automatic control laboratory, ETH Zürich, Switzerland. Email: \{mbianch,dorfler\}@ethz.ch. This work was supported by ETH Zürich funds. 
% \textit{Corresponding Author}: Duong Thuy Anh Nguyen.}
}
}

\maketitle

\begin{abstract} 
In this paper, we address the challenge of \gls{NE} seeking in non-cooperative convex games with partial\mattiasay{-decision } information. We propose a distributed \mattiasay{algorithm,
%consensus-based projected-gradient algorithm, 
where each agent refines its strategy through projected-gradient steps and an averaging procedure}. Each agent uses estimates of competitors' actions obtained solely from local neighbor interactions, in a directed communication network. 
%To control the update speed, an averaging parameter is incorporated into the update step, balancing information from neighbors with the agent's current estimate.
Unlike previous approaches that rely on \mattiasay{(strong) } monotonicity assumptions, this work establishes the convergence towards a NE under a diagonal dominance property
%of the Jacobian matrix
of the pseudo-gradient mapping, \mattiasay{that can be checked locally by the agents. Further, this } condition is \mattiasay{physically interpretable } and of relevance for many applications, as it suggests that an agent's objective function is primarily influenced by its individual strategic decisions, rather than by the actions of its competitors. \mattiasay{In virtue of a novel block-infinity norm convergence argument}, we provide explicit bounds for constant step-size that are independent of the communication structure, \mattiasay{and can be computed in a totally decentralized way}. Numerical simulations on an optical network's power control problem validate the algorithm's effectiveness.
\end{abstract}

\printnomenclature
\allowdisplaybreaks

%%%%%%%%%%%%%%%%%%%%%%%%%%%%%%%%%%%%%%%%%%%%%%%%%%%%%%%%%%%%%%%%%%%%%%%%%%%%%%%%
\section{Introduction} \label{intro}
 \glsfirst{NE} computation holds pivotal significance within the realm of multi-agent decision learning, particularly in the context of non-cooperative games. In such games, individual agents pursue their unique, rival-dependent objectives devoid of any inter-agent coordination. 
 %with the selected strategy often influenced by agents' expectations of others' behaviors. 
\mattiasay{Applications include }power control and resource allocation in wireless and communication networks \cite{Pan2007,Naik2010}, distribution grids \cite{Belgioioso_OperationallySafe_TSG22}, cognitive radio systems \cite{Scutari_ComplexGames_TIT2014}, 
% smart grids \cite{BasharSG}, 
crowdsourcing \cite{wiopt23}, and formation control \cite{Simaan_AUT_2019}. Significant research efforts have been dedicated to developing effective techniques for computing a \gls{NE}, including decomposition methods like Jacobi or Gauss-Seidel-type approaches \cite{Facchinei_Kanzow_2010,BorgensKanzow2018}, variational inequality-based approach \cite{Facchinei_Pang_Variational_2009,Facchinei_Kanzow_2010}, best-response strategy \cite{Huang_Hu_arXiv_2021,Bianchi_GNEPPP_AUT_2022}, fictitious play \cite{Shamma_Arslan_2005,SwensonKarXavier_FictitiousPlay_2015} and gradient-based methods \cite{Belgioioso_Grammatico_Preconditioned_ECC_2018,Bianchi_Directed_CDC_2020,Bianchi_TV_LCSS_2021,Tatarenko2021,nguyen2022distributed,Nguyen2023AccGame}. 

Classical computational settings typically assume that each agent or a centralized coordinator possesses comprehensive information regarding local objective functions, decision sets and actions of its competitors~\cite{Facchinei_Pang_Variational_2009,Facchinei_Kanzow_2010,Belgioioso_Grammatico_Preconditioned_ECC_2018}. However, this presumption often becomes impractical when applied to real-world systems, especially in games played over large-scale networks. 
Given the proliferation of engineering networked applications, exemplified by vehicle-to-vehicle, peer-to-peer, ad-hoc, and smart-grid networks, there has been increasing interest in developing distributed protocols to find \an{a} \gls{NE} in the ``partial-decision information'' scenario, where agents have access only to the information exchanged with neighbors over a communication network \cite{Bianchi_Directed_CDC_2020,Bianchi_TV_LCSS_2021,Tatarenko2021,nguyen2022distributed,Nguyen2023AccGame}. Such algorithms not only respect privacy concerns, but also offer robustness and scalability. 

In this partial decision information scenario,  dealing with directed communication is one of the main challenges. In fact, early works have predominantly focused on undirected networks and often assume doubly-stochastic weights. Various approaches have been proposed, based on consensus-based methods, such as fictitious play in congestion games \cite{SwensonKarXavier_FictitiousPlay_2015}, projected-gradient algorithms for aggregative games \cite{Koshal2016} and generally coupled cost games \cite{Tatarenko2021}, inexact ADMM-type algorithms \cite{Salehisadaghiani_Wei_Pavel_AUT_2019}, operator splitting approach \cite{Pavel_GNE_TAC_2020} and proximal best-response strategies \cite{Bianchi_GNEPPP_AUT_2022}. \mattiasay{However,  in 
many scenarios, considering directed communication is imperative, in order to model unilateral transmission capability, e.g., in 
wireless systems with heterogeneous power levels or bandwidth.  In this case, even  considering doubly-stochastic weights 
(or weight-balanced weights, as in~\cite{Bianchi_TV_LCSS_2021})  is not a viable option, \an{since} assigning such weights over directed graphs is complex and computationally demanding. }

\mattiasay{For this reason, some studies \an{have} focused on the case of direct networks and row-stochastic matrices, which are straightforward to construct by } allowing each agent to assign weights to incoming information locally. However, row-stochasticity introduces technical complications, as it leads to the loss of various properties associated with doubly-stochastic matrices. Reference \cite{SalehisadaghianiPavel2017_nondoublystochastic} highlights these issues, leading to the development of an asynchronous gossip algorithm. 
% assuming each agent can update estimates from all agents impacting its cost function.
Reference \cite{Bianchi_Directed_CDC_2020} directly addresses row-stochastic weights, but it requires complete knowledge of the communication structure to compute the Perron-Frobenius eigenvector of the adjacency matrix; \mattiasay{this condition is then removed in \cite{Bianchi_minG_TAC_2022}}. In\cite{nguyen2022distributed,Nguyen2023AccGame}, these assumptions are further relaxed \mattiasay{for time-varying networks, } and a condition is introduced to address the common issue of monotonicity loss with weighted norms. 
% In the subsequent work \cite{Nguyen2023AccGame}, an accelerated algorithm is proposed for the unconstrained game, leveraging heavy-ball momentum. 
In a recent study \cite{GadjovTCSN2023}, a resilient algorithm is proposed utilizing an observation graph \mattiasay{and row-stochastic mixing matrices.}
% assuming unconstrained action sets and direct observation of actions from truthful agents, making it resistant to tampering by adversarial agents.

For all the all of the algorithms mentioned above, the monotonicity of the pseudo-gradient mapping $F$ stands as a standard and widely accepted assumption for ensuring convergence. In particular, the strong monotonicity of $F$ is sufficient to guarantee the existence and uniqueness of the \gls{NE}, and it is frequently adopted in algorithms with fixed step-sizes, as seen in references such as  \cite{Belgioioso_Grammatico_Preconditioned_ECC_2018,Salehisadaghiani_Wei_Pavel_AUT_2019,Tatarenko2021,Bianchi_Directed_CDC_2020,Bianchi_TV_LCSS_2021,Huang_Hu_arXiv_2021,nguyen2022distributed,Nguyen2023AccGame,GadjovTCSN2023}. This assumption is alternatively replaced by strict monotonicity, particularly when considering vanishing step-sizes \cite{Koshal2016,SalehisadaghianiPavel2017_nondoublystochastic}.
% In broad terms, $F$ being monotone is one of the weakest conditions under which global convergence can be proved \cite{FacchineiPang}.
Nevertheless, monotonicity \mattiasay{can be } a highly restrictive requirement \cite{Facchinei_Kanzow_2010}. Specifically, it is well known that a continuously differentiable mapping $F$ is  monotone \mattiasay{if the symmetric part } of its Jacobian is positive-semidefinite \ella{\cite[Proposition 12.3]{Rockafellar_Wets_1998}}. This imposes a strong assumption on the structure and relations among the objective functions of all agents:
\mattiasay{in particular, monotonicity is virtually impossible to verify in a distributed way. }

\textbf{Contributions.} Motivated by the above, we study \gls{NE}  seeking in non-cooperative convex games over directed graphs. 
In contrast to existing works that rely on the conventional assumption of strong monotonicity for the pseudo-gradient mapping \cite{Belgioioso_Grammatico_Preconditioned_ECC_2018,Salehisadaghiani_Wei_Pavel_AUT_2019,Tatarenko2021,Bianchi_Directed_CDC_2020,Bianchi_TV_LCSS_2021,nguyen2022distributed,Nguyen2023AccGame,GadjovTCSN2023}, we instead assume a diagonal  dominance \mattiasay{condition, similar to one considered in \cite{nguyen2022distributed} and \cite{Scutari_ComplexGames_TIT2014}. This condition does not imply monotonicity and can be checked locally by the agents. If the mapping $F$ is differentiable, it corresponds to block-diagonal domincance of its Jacobian; }
intuitively, this suggests that each agent has more influence over their own objective function compared to the other agents. Indeed, the diagonal dominance assumption finds applicability in   \gls{NE} problems arising in various engineering domains, including the power control problem in optical networks \cite{Pan2007}, cognitive radio networks \cite{Scutari_ComplexGames_TIT2014}, interference coupled wireless systems \cite{Naik2010}, Nash-Cournot game \cite{Kolstad1991}, newsvendor pricing game \cite{Chen2004}, investment portfolio selection problem \cite{Lei2017} and banking risk management \cite{Bichuch2020}. 

Under this condition, we study a novel 
distributed consensus-based projected-gradient algorithm for \gls{NE} seeking under partial-decision information scenario. In this algorithm, each agent performs a gradient step to minimize its own cost function using estimates of competitors' actions, which are derived solely from information exchanged with neighbors through a \mattiasay{row stochastic directed } communication network. We introduce an averaging parameter in the update step to control the speed at which each agent adjusts its strategy, balancing the influence of information received from neighbors against its current estimate. 
% of the mapping's Jacobian and establish the convergence of the algorithm to the \gls{NE} for a strongly connected directed communication graph. 
Significantly, \mattiasay{for the first time in the partial-decision information scenario, } we provide explicit bounds for the constant step-sizes  that possess the desirable property of being independent of the communication graph's connectivity structure, allowing for local computation based on the properties of each agent's private objective function. The performance of the proposed algorithm is demonstrated through numerical simulations on a power control problem in optical networks as described in \cite{Pan2007}.

% This paper is structured as follows: Section~\ref{sec:formu} presents the problem formulation, Section~\ref{sec:algo}introduces the distributed algorithm and provides convergence analysis, Section~\ref{sec:simulation} numerically evaluates the performance of the proposed algorithm, and Section~\ref{sec:conc} concludes with key points.

\textbf{Notations.} 
All vectors are column vectors unless otherwise stated and we consider a discrete time model where the time index is denoted by $k$. We write $u^\top$ for the transpose of a vector $u\in \re^n$. We use $\0$ and $\1$ to denote the vector with all entries equal to $0$ and $1$, respectively.  The $i$-th entry of a vector $u$ is denoted by $u_i$. Given a vector $u$, $\diag(u)$ denotes the diagonal matrix whose diagonal entries correspond to the entries of $u$. We use $[A]_{ij}$ \an{or $a_{ij}$} to denote the $ij$-th entry of a matrix $A$. A matrix $A$ is nonnegative if all its entries are nonnegative. A nonnegative matrix $A\in\mathbb{R}^{n\times n}$ is row-stochastic if $A\mathbf{1}=\mathbf{1}$. The identity matrix is denoted by $\mbi$ and $ \| \cdot \| $ denotes the standard Euclidean vector/matrix norm.

% Let  $E = \re^n$.
% or the space of real matrices $E = \re^{n\times n}$.
\an{
We use $\proj_\Omega (u)$ to denote the Euclidean projection of $u \in \re^n$ to a set $\Omega \subseteq \re^n$. A mapping $g:\re^n\to \re^n$ is said to be \textit{strongly monotone} on a set $Q\subseteq \re^n$ with the constant $\mu>0$, if $\langle g(x)-g(y), u - v \rangle\ge\mu\|x - y\|^2$ for any $x,y\in Q$; it is said to be \textit{Lipschitz continuous} on a set $Q\subseteq \re^n$ with the constant $L>0$, if $\|g(x)-g(y)\|\le L\|x - y\|$ for any $x,y\in Q$; it is said to be nonexpansive with respect to a norm $\|\cdot\|_v$ if $\|g(x)-g(y)\|_v\leq \|x - y\|_v$, for all $x,y\in \re^n$.
}

Let $\mc{I}=\{1,\ldots,N\}$ with an integer $N\ge 1$. Given a directed graph $\G=(\mc{I},\mc{E})$, specified by the set of edges $\mc{E}\subseteq \mc{I}\times\mc{I}$ of ordered pairs of nodes, the in-neighbor and out-neighbor set for every agent $i$ are defined, as follows:
\[\cNini=\{j\!\in\!\mc{I}|(j,i)\!\in\!\mc{E}\} \!\!\!\text{and}\!\!\! \cNouti=\{\ell\!\in\!\mc{I}|(i,\ell)\!\in\!\mc{E}\}.\] 
A directed graph $\G$ is {\it strongly connected} if there is a directed path from any node to all other nodes in $\G$.

We recall the following fundamental result
\mattiasay{
\begin{lemma}[{Krasnoselski-Mann iteration, \cite[Theorem 3]{borwein_reich_shafrir_1992}}] \label{lem:KM}
    Assume that $g:E\rightarrow E$ is nonexpansive with respect to a norm $\| \cdot\|_v$ and that $\operatorname{fix} (g) = \{ x\in E \mid g(x) =x\} \neq \varnothing$. Then, for any scalar $\gamma \in (0,1)$, the iteration
    \begin{align}
        x^{k+1} = (1-\gamma)x^k +  \gamma g (x^k), \quad \forall k \in \mathbb{N}
    \end{align}
    converges to a point   $x^\star \in \operatorname{fix}(g)$, for any  $x^0 \in E$. 
\end{lemma}
}

\section{Problem Formulation} \label{sec:formu}
\subsection{The Game}
We consider a non-cooperative game with a set of  agents, $ \mc I\coloneqq \{ 1,\ldots,N \}$, where each agent $i\in \mc{I}$ shall choose its decision variable  $x_i \in \Omega_i \subset \R^{n_i}$, for $i\in \mc I$. 
\an{Let $x \coloneqq  \col( (x_i)_{i \in \mc I})$ 
denote the stacked vector of all the agents' decisions, which belongs to the joint decision set $\Omega \coloneqq \Omega_1\times \dots \times\Omega_N $, and let $\textstyle n\coloneqq \sum_{i=1}^N n_i$.}
The goal of each agent $i \in \mc I$ is to minimize its objective function $J_i(x_i,x_{-i})$, that depends on both the local variable $x_i$ and the decision variables of the other agents $x_{-i}\coloneqq  \col( (x_j)_{j\in \mc I\backslash \{ i \} })$. $x_{-i} \in \Omega_{-i}\subset\re^{n-n_i}$. The game, denoted as $\Gamma=(\mc I,\{J_i\},\{\Omega_i\})$, is characterized by a set of interdependent optimization problems given as follows:
	\begin{align} \label{eq:game}
	(\forall i \in \mathcal{I})
	\  \min_{ y_i \in \Omega_i}   \; J_i(y_i,x_{-i})
	\end{align}
The technical problem we consider here is the computation of a \gls{NE} of the game $\Gamma$, namely a set of decisions that simultaneously solve all the optimization problems in  \eqref{eq:game}.

\begin{definition}[Nash equilibrium]
A collective strategy $x^{\star}=\col\left((x_{i}^{\star}\right)_{i \in \mathcal{I}})$ is a  Nash equilibrium if, for all $i \in \mc{I}$,
	$
			% \quad x_{i}^{\star} \in \underset{y_i}{\argmin} \, J_{i}\left(y_{i}, x_{-i}^{\star}\right) \text { s.t. } (y_{i},x^{\star}_{-i})\in \mc{X}.
		J_{i}\left(x_i^{\star}, x_{-i}^{\star}\right)\leq \inf \{J_{i}\left(y_{i}, x_{-i}^{\star}\right) \mid (y_i,x_{-i}^{\star})\in\Omega \}.$
	\end{definition}
We postulate throughout standard regularity and convexity assumptions.
    \begin{standing}[Convexity]\label{asm:Convexity} 
        For each $i\in \mathcal{I}$, the action set $\Omega_i$ is non-empty, closed and convex. The objective function $J_i$ is continuous and $J_{i}\left(\cdot, x_{-i}\right)$ is also convex and continuously differentiable for every $x_{-i}\in \Omega_{-i}$. The set of Nash equilibria, denoted as $\Omega^*$, is nonempty.
    \end{standing}

Under \cref{asm:Convexity}, a \gls{NE} of the game can be alternatively characterized using the first-order optimality conditions. Specifically, $x^*\in \Omega$ is a \gls{NE} if and only if, for all $i\in\mc{I}$, the following condition holds:
\beqn
\langle \nabla _{\!x_i} J_i(x_i^*,x_{-i}^*),x_i-x_i^*\rangle\ge 0, \quad  \forall x_i \in \Omega_i.
\eeqn
The precedent relation is equivalent to: 
\beqn
\label{eq-agent-fixed-point}
x_i^*=\proj_{\Omega_i}[x_i^*-\a_i \nabla_{\!x_i} J_i(x_i^*,x_{-i}^*)], \quad \forall i\in\mc{I},
\eeqn
where $\a_i>0$ is an arbitrary scalar.

% Since we focus on gradient-type methods with fixed step-sizes, we also assume smoothness of the cost functions. 
\an{Next, we define the pseudo-gradient 
\begin{align}
	\label{eq:pseudo-gradient}
	F(x)\coloneqq \col\left( (\nabla _{\!x_i} J_i(x_i,x_{-i}))_{i\in\mc{I}}\right),
\end{align}
for which we make the following assumption.}
\begin{standing}[Lipschitz continuity] \label{asm:Lipschitz}
    The pseudo-gradient operator $F$ is $\ell$-Lipschitz continuous, for some $\ell >0$.
\end{standing}

This assumptions implies that, for every $i\in\mc{I}$, each mapping $\nabla_{x_i} J_i$ is $\ell$-Lipschitz considering the entire argument. Additionally, we denote by $\ell_{i,j}$ the Lipschitz constant of $\nabla_{x_i} J_i(x)$ with respect to $x_j \in \Omega _j$, for all $i,j\in\mc{I}$. 
% Furthermore, the previous assumptions implies that $F$ is almost everywhere differentiable, with jacobian denoted by $DF(x) \in \R^{n\times n}$, whenever $F$ is differentiable at $x$. 

\subsection{The Communication}
We consider the so-called partial-decision information scenario for the game $\Gamma=(\mc I,\{J_i\},\{\Omega_i\})$, in which there is no central coordinator. In this scenario, agents are limited to exchanging information exclusively through peer-to-peer communication, facilitated by a directed communication graph denoted as 
\an{$\mathbb{G} =(\mc{I},\mc{E})$. A weight matrix $W \in \mathbb{R}^{N \times N}$ is associated with this graph, such that $w_{ij}>0$ \an{whenever} $(j,i)$ belongs to the edge set $\mc{E}$,} signifying that agent $i$ can receive information from agent $j$.

\begin{standing}
 [Communication]\label{asm:graph}
The graph $\mathbb{G}$ is strongly connected. The weight matrix $W$ satisfies:
\begin{itemize}
    \item \emph{Row stochasticity}: $ W \1 =  \1$;
    \item \emph{Self-loops}: $w_{ii} >0 $ for all $i\in \mc{I}$;
    % \item \emph{Nonvanishing weights}: There exists a scalar $ \epsilon >0$ such that $w_{ij} >\epsilon$ whenever $w_{ij}>0 $. 
    \end{itemize}
\end{standing}

\subsection{Partial-Decision Information Notations} \label{sec:notation}
To remedy the lack of global knowledge, we let each agent $i\in \mc{I}$ maintain an estimate $\x_{i,j}  \in \R^{n_j}$ of the actions of every other agent $j$.
We  define $\x_{i,i} \coloneqq x_i$,
\begin{align}
\x_{i} & \coloneqq \col((\x_{i,j})_{j\in\mc{I}})\in\R^{n}, \label{eqn:ximat}
\\
\x& \coloneqq \col((\x_i)_{i\in\mc{I}})\in\R^{Nn}. \label{eqn:xmat}
\end{align}
Furthermore, let us define, as in \cite[Eq.~13]{Pavel_GNE_TAC_2020}, for all $i \in \mc{I}$,
	% \begin{subequations}
		\begin{align}
		\mathcal{R}_{i}\coloneqq & \begin{bmatrix}{{\0}_{n_{i} \times n_{<i}}} & {\mbi_{n_{i}}} & {\0_{n_{i} \times n_{>i}}}\end{bmatrix} \in \R^{n_i \times n },
		% \\ 
		% \mathcal{S}_{i}\coloneqq  &\begin{bmatrix}{\mbi_{n_{<i}}} & {\0_{n_{<i} \times n_{i}}} & {\0_{n_{<i} \times n_{>i}}} \\ {\0_{n_{>i} \times n_{<i}}} & {\0_{n_{>i} \times n_{i}}} & {\mbi_{n_{>i}}}\end{bmatrix} \in \R^{n_{-i}\times n }\hspace{-2pt}
		\end{align}
	% \end{subequations}
where $n_{<i}\coloneqq \sum_{j<i,j \in \mathcal{I}} n_{j}$ and $n_{>i}\coloneqq \sum_{j>i, j \in \mathcal{I}} n_{j}$. In simple terms, $\mathcal R _i$ selects the $i$-th component, which is of dimension $n_i$, from a vector of size $n$. Thus, 
\[\mathcal{R}_{i} \x_{i}=\x_{i,i}=x_i.\]
% \[\mathcal{R}_{i} \x_{i}=\x_{i,i}=x_i \text{ and } \mathcal{S}_{i} \x_{i}=\x_{i,-i}.\]
Let $\mathcal{R}\coloneqq \diag\left((\mathcal{R}_{i})_{i \in \mathcal{I}}\right)$. \an{Hence, the vector $x$ of stacked agents' decisions satisfies} $x=\mathcal{R} \x$.
% and the stacked estimates, denoted as $\z = \col((\x_{i,-i})_{i \in \mathcal{I}})$, can be represented as $\z=\mathcal{S}{\x} \in \R^{(N-1) n}$. We further have,
	% % \begin{subequations}
	% % 	\label{eq:RSproperties}
	% % 	\begin{gather}
	% % 	\label{eq:RSproperties:a}
	% % 	\mathcal{R}^\top \mathcal{R}+\mathcal{S}^\top \mathcal{S}=\mbi_{N n}
	% % 	\\
	% % 	\label{eq:RSproperties:b}
	% % 	{\mathcal{R S}^\top=\0_{n}, \quad  \mathcal{S R}^\top=\0_{N n-n}}
	% % 	\\
	% % 	\label{eq:RSproperties:c}
	% % 	{\mathcal{R} \mathcal{R}^\top=\mbi_{n}, \quad \mathcal{S} \mathcal{S}^\top=\mbi_{N n-n}}
	% % 	\\
	% % 	\label{eq:RSproperties:d}
	% % 	\x=\mathcal{R}^\top x+\mathcal{S}^\top \v.
	% % 	\end{gather}
	% % \end{subequations}
	% $\mathcal{R}^\top \mathcal{R}+\mathcal{S}^\top \mathcal{S}=\mbi_{N n}$ and $\x=\mathcal{R}^\top x+\mathcal{S}^\top \z$.
 
	In the partial-decision information setting, agents depart from the conventional approach of evaluating gradient at actual decisions, $\nabla_{\!x_i}\! J_i(x_i,x_{-i})$, and instead each agent computes the gradient based on local estimates, namely $\nabla_{\!x_{i}}\! J_{i}\left(x_{i}, \x_{i,-i}\right)$. To formalize this notion, we introduce the \textit{extended pseudo-gradient} mapping, $\bs{F}\!:\R^{Nn}\rightarrow \R^n$, defined as:
	\begin{align}
	\label{eq:extended_pseudo-gradient}
	\bs{F}(\x)\coloneqq \col\left((\nabla_{\!x_{i}} J_{i}\left(x_{i}, \x_{i,-i}\right))_{i \in \mathcal{I}}\right).
	\end{align}
    This mapping extends the pseudo-gradient $F$ defined in \eqref{eq:pseudo-gradient} to the augmented space of decisions and estimates, denoted as $\bs{\Omega}\coloneqq \{\x\in\R^{nN}\mid \mc{R}\x\in \Omega\}$.

    % At times, it will be useful to consider action-wise (instead of agent-wise) quantities, which we denote with a tilde, as in
    % \begin{align}
    %     \tilde{\x}_i & \coloneqq \col((\x_{j,i})_{j\in \mc{I}}) \in \R^{N n_i}\\
    %     \tilde{\x} & \coloneqq \col((\tilde{\x}_i)_{\i});
    % \end{align}
    % note that $\tilde{\x}_i$ collects the estimate of $x_i$ maintained by different agents. Let $\Pmat \in \R^{Nn\times Nn}$ denote the permutation matrix such that $\Pmat \x = \tilde{\x}$. This matrix is instrumental in defining all agent-wise quantities; for instance, $\tilde{\bs{\Omega}} \coloneqq \mathbf{P} \bs\Omega$.
    
\subsection{Alternative assumption}
In the domain of games involving partial-decision information, a standard assumption is that the game mapping $F$ exhibits strong monotonicity property. This assumption implies that the functions $J_i$ are strongly convex with respect to the local variables and ensures the existence of a unique \gls{NE} for the game $\Gamma=(\mc I,\{J_i\},\{\Omega_i\})$ in \eqref{eq:game}. Deviating from this conventional assumption, our work focuses on algorithms that  converge to a \gls{NE} under the following assumption: 

% \begin{assumption}[Diagonal dominance, unconstrained, fixed graph] ~
%     \begin{enumerate}[noitemsep,nolistsep]
%     \item For each $x \in \R^n$, $DF(x)$ is row diagonally dominant, i.e.,
%     \begin{align*}
%     (\forall l \in \{1,2,\dots,n\})  \qquad [DF(x)]_{l,l} \geq  \textstyle \sum_{ m \in \{1,2,\dots,n\} \backslash \{l\}} \left| [DF(x)]_{l,m}  \right|
%     \end{align*}
%     \item The communication graph is time invariant (i.e., $\mathbb{G}^k = \mathbb{G}$ for all $\k$) and strongly connected.
% \end{enumerate}
% \end{assumption}

\begin{assumption}[Block-diagonal dominance] \label{asm:DD_FG} ~
For all $\i$, $J_i(\cdot,x_{-i})$ is $\mu_i$--strongly convex for any $x_{-i} \in \re^{n-n_i}$, where $\mu_i>0$. Furthermore,
    \begin{align}\label{eq:diagonaldominance}
        \mu_i \geq \textstyle \sum_{j\in {\mc{I} \backslash \{i\} }} \ell_{i,j}.
    \end{align}
\end{assumption}

% \begin{assumption}[Convergence conditions] 
% \label{asm:alternative}
% \begin{enumerate}[label=(\roman*)]
%     \item \label{asm:alternative:i}
%  \emph{Fixed graph:} The communication graph is time invariant (i.e., $\mathbb{G}^k = \mathbb{G}$ for all $\k$) and strongly connected;
%     \item \label{asm:alternative:ii} \emph{Unconstrained game:} $\Omega = \R^n$;
%     \item \label{asm:alternative:iii} \emph{Strong monotonicity:} The pseudo-gradient mapping $F$ is $\rho$-strongly monotone, for some $\rho > 0$;
%     \item \label{asm:alternative:iv} \emph{Monotonicity in $\infty$-norm:} There exists $\rho_\infty \geq 0$ such that, for every $x$ where $F$ is differentiable, it holds that $ -\mu_\infty (-DF(x)) \geq \rho_\infty$. \hfill $\square$
% \end{enumerate}
% \end{assumption}

\begin{remark}\label{remark-diagonal-dominance-1} If $F$ is differentiable,
\cref{asm:DD_FG} pertains to the row-block-diagonal dominance of the Jacobian matrix associated with the mapping $F$. Essentially, this property stipulates that for each agent $i\in\mc{I}$, their objective function $J_i$ is primarily influenced by their own decision variables or choice of strategies, rather than being significantly affected by the collective strategies of their competitors. This type of dominance property is intuitive and appears in many applications \cite{Pan2007,Scutari_ComplexGames_TIT2014,Naik2010,Kolstad1991,Chen2004,Lei2017,Bichuch2020}. 
\end{remark}

\begin{remark}\label{remark-diagonal-dominance-2}
\cref{asm:DD_FG} aligns with those assumed in recent studies on distributed \gls{NE} seeking algorithms, such as \cite[Proposition 7]{Scutari_ComplexGames_TIT2014}, \cite[Remark 3]{Tatarenko2021}, and the condition $\beta_\alpha > 0$ in \cite{nguyen2022distributed}.
Notably, \cref{asm:DD_FG} is considerably weaker than the assumptions put forth in the recent literature \mattiasay{for games in partial-decision information }\cite{nguyen2022distributed}, where strong monotonicity is assumed along with conditions stricter than \cref{asm:DD_FG}. \mattiasay{Furthermore, \cref{asm:DD_FG} does not requires uniqueness of the \gls{NE}, differently from strong monotonicity, or,  e.g., the condition in \cite[Th. ~10]{Scutari_ComplexGames_TIT2014}.}
\end{remark}
% consistently played a central role within the contraction approach for achieving convergence in distributed algorithms.

% \cref{asm:DD_FG}, in general, does not necessarily imply strong monotonicity. To illustrate, consider the linear operator:
% \begin{align}
%     F(x) = \begin{bmatrix}
%         1 & -1 \\
%         -1 &1
%     \end{bmatrix}x.
% \end{align}
% This operator satisfies \cref{asm:DD_FG} but is not strongly monotone and is not even monotone. It is monotone with respect to a weighted norm, however, working with weighted norms can be problematic in distributed setups, especially when dealing with multiple operators that may not be monotone in the same norm or when the weighted norm is unknown.

\begin{remark}\label{remark-diagonal-dominance-3}
\mattiasay{
If \eqref{eq:diagonaldominance}
holds strictly, then the game in \eqref{eq:game} has a unique \gls{NE} (this can be inferred using weighted norms by \cite[Theorem 4.1]{Gabay1980}, \cite[Theorem 4]{Chen2004}). Nonetheless, even in this case the game mapping may be not (strongly) monotone. For example consider the game corresponding to 
\begin{align}
    F(x) = \begin{bmatrix}
        1 & -0.9 \\
        -9 &  10
    \end{bmatrix}x,
\end{align}
for which \eqref{eq:diagonaldominance} holds strictly: the operator $F$ is not monotone.}
\end{remark}

% Gabay and Moulin's result holds while making no assumption on compactness. However, notice that the Gabay and Moulin assumptions become quite demanding, if they are applied to Cournot oligopoly games when the number of agents increases.

\section{Distributed Nash Equilibrium Seeking} \label{sec:algo} 
Consider the game $\Gamma=(\mc{I},{J_i},{\Omega_i})$ where agents interact through a directed communication graph $\G=(\mc{I},\mc{E})$. Given the constraints on agents' access to others' actions in game $\Gamma$, we propose a fully-distributed algorithm that respects the information access as dictated by the communication graph $\G$. The approach is outlined in Algorithm 1.

At each time step $k$, every agent $i$ shares its estimation $\x_i^k$ with its out-neighbors $\ell \in \cNouti$, while receiving estimates $\x_j^k$ from its in-neighbors $j \in \cNini$. Subsequently, agents collectively aggregate the current estimates of their neighbors by computing a weighted average and initiate a local gradient step. This step is governed by the step-sizes and aligns with the gradient of their respective cost functions evaluated at the most recent local estimates. Let $\bs W  \coloneqq W \otimes \mbi_N$. The local updates above can be represented in a compact form, as follows:
\begin{align}
    \label{eq:algo:CASE1}
    \x^{k+1} = \gamma \x^k+ (1-\gamma)\proj_{\bs{\Omega}}\!\left (W\x^k \!- \mc{R}^\top \!\Lambda\Fbs (\bs W\x^k)\right)\!,\!\!
\end{align}
where $\Lambda = \diag((\alpha_i\mbi_{n_i} )_{\i}) \succ 0$ is the matrix of step-sizes and $\gamma \in (0,1)$ is an averaging parameter. 

\begin{remark}
It is important to note that for the algorithm to maintain its iterative nature and provide meaningful updates, $\gamma$ needs to be strictly less than $1$. If $\gamma=1$, the equation \eqref{eq:algo:CASE1} becomes $\x^{k+1} = \x^k$ for all $k\geq 0$, the agent essentially maintains its current estimate unchanged throughout the iterations, which results in no updates and a lack of algorithmic progress. 
\end{remark}

\begin{remark}\label{rem-AveragingParam}
The choice of $\gamma$ influences how quickly an agent adjusts its strategy based on the information it receives from neighbors. When $\gamma$ is close to $0$, the agent gives more weight to the received information, potentially leading to faster convergence to a consensus or equilibrium. In the case where $\gamma=0$, the algorithm is proposed, and its convergence analysis is presented in \cite{nguyen2022distributed}. Conversely, as $\gamma$ approaches $1$, the agent places more emphasis on its current estimate, which can slow down convergence but might be advantageous in avoiding oscillations or overshooting.
\end{remark}

The update in \eqref{eq:algo:CASE1} is equivalent to each agent $i \in \mc{I}$ updates its local estimate of the joint action as follows: 
\begin{subequations}
\label{eq:algo:DD_FG} 
\begin{align}
\!\!\hat \x_i^{k+1} &= \textstyle\sum_{j\in \I} w_{ij} \x_{j}^k
\\ 
\!\!\x_{i,-i}^{k+1} & = \gamma \x_{i,-i}^k+ (1-\gamma)\hat \x_{i,-i}^{k+1}
\\
\!\!x_i^{k+1} &=\gamma x_i^k \!+\! (1-\gamma) \proj_{\Omega_i} \!\!\left ( \!\hat \x_{i,i}^k\!-\! \alpha_i \nabla_{x_i} J_i\!\left(\hat \x_i^{k+1}\right) \!\right)\!.\!\!
\end{align}
\end{subequations}

\begin{table}[t!]
\centering \normalsize
\vspace{0.25cm}
    \begin{tabular}{l}
    \hline
    \multicolumn{1}{c}{\textbf{Algorithm 1: Distributed \gls{NE} Seeking}}\\
    \hline
    Every agent $i\in\mc{I}$ selects a local step-size $\a_i>0$,\\ chooses the weights $[W_k]_{ij},j\in\mc{I}$, and initializes\\ with arbitrary vectors $\x_{i,-i}^0\in\re^{n-n_i},x_i^{0}\in\Omega_i$.\\
    \textbf{for} $k=0,1,\ldots,$ every agent $i\in\mc{I}$ does the following:\!\!\\
    \emph{  } Receives $\x_j^k$ from in-neighbors $j\in\cNini$;\\
    \emph{  } Sends $\x_i^k$ to  out-neighbors $\ell\in\cNouti$;\\
    \emph{ } Updates the action $x_i^{k+1}$ and estimates $\x_{i,-i}^{k+1}$ by \eqref{eq:algo:DD_FG};\\
    \textbf{end for}\\
    \hline
    \end{tabular}
    \vspace{-0.5cm}
\end{table}

% \section{Convergence of \textbf{DNE-HB}}\label{sec:conv_results}
It clearly holds that the fixed points of \eqref{eq:algo:CASE1} coincide with the vectors $\x^* = \1 \otimes x^*$, where $x^*$ is a \gls{NE} as specified \eqref{eq-agent-fixed-point}. In the following, we proceed to demonstrate the convergence of the dynamics when sufficiently small step-sizes are employed.

\begin{theorem}[Convergence of \eqref{eq:algo:DD_FG}] \label{theo:conv}
Let \cref{asm:Convexity}, \cref{asm:Lipschitz}, \cref{asm:graph} and \cref{asm:DD_FG} hold. Let the step-sizes satisfy, for all $i\in\mc{I}$:
\begin{align}
    0< \alpha_i < \frac{1}{\ell_{i,i}}.
\end{align}
Then, for any initial condition $\x^0$ and any $\gamma \in (0,1)$, the sequence $(\x^k)_\k$ generated by \eqref{eq:algo:CASE1} converges to a point $\x^* = \1\otimes x^*$, where $x^* \in \Omega^*$. 
\end{theorem}

\begin{proof}
Let us rewrite the update in \eqref{eq:algo:CASE1} as 
\begin{align}\label{eq:KM}
    \x^{k+1} = \gamma \x^k +(1-\gamma) \mc{A} (\x^k)
\end{align}
where the operator $\mc{A}$ is the composition of three operators
\begin{align}\label{eq:Aop}
    \mc{A}  \coloneqq \proj_{\bs{\Omega}} \circ (\operatorname{Id} - \Rmc^\top \Lambda \bs F) \circ \bs W.
\end{align}
Here, $\operatorname{Id}$ is the identity operator.

In the following, we show that the operator $\mc{A}$ is nonexpansive (i.e., $1$--Lipschitz) with respect to a \an{suitably chosen} norm. Then, the convergence result follows because  \eqref{eq:KM} corresponds to the standard Krasnoselskii--Mann iteration in \cref{lem:KM} applied to the operator $\mc{A}$.

In particular, for any $\x\in\R^{Nn}$ defined as in \eqref{eqn:xmat}, we consider the mixed norm defined as follows
\begin{align}
    \| \x \|_v \coloneqq \| \x \|_{2,\infty} = \left \| \begin{matrix}
        \| \x_{1,1} \|
        \\
        \| \x_{1,2} \|
        \\
        \vdots 
        \\
        \| \x_{N,N} \|
    \end{matrix}\right\|_{\infty}.
\end{align}

The following lemmas show the nonexpansiveness property of each of the three operators in \eqref{eq:Aop}.

 \begin{lemma}
 The operator $\bs{W}$ is nonexpansive with respect to the mixed norm $\|\cdot\|_v$.
 \end{lemma}
 \begin{proof}
    For any $\x\in\R^{Nn}$ defined as in \eqref{eqn:xmat}, we have 
    \begin{align*}
        \| \bs W \x \|_v &= \max_{\i} \max_{j\in\mc{I}} \sum_{l\in\mc{I}} w_{il} \x_{l,j} 
        \\
        & \leq  \max_{\i} \max_{j\in\mc{I}} \max_{l\in\mc{I}} \| \x_{l,j} \| = \|\x\|_v,
    \end{align*}
     where the inequality follows by the row-stochasticity property of the weight matrix $W$ (\cref{asm:graph}). \hfill
 \end{proof}

 \begin{lemma}
     The projection operator $\proj_{\bs{\Omega}}$ is nonexpansive with respect to the mixed norm $\|\cdot\|_v$.
 \end{lemma}
 \begin{proof}
     The proof follows directly from the nonexpansiveness property of each projection $\proj_{\Omega_i}$ in the Euclidean norm \cite[Prop.~4.16]{Bauschke_2017} and definition of $\|\cdot \|_v$. \hfill
 \end{proof}

 \begin{lemma} \label{lem:OperatorIdRF}
     The operator $ \operatorname{Id}- \mc{R}^\top \Lambda\bs{F}$ is nonexpansive with respect to the mixed norm $\|\cdot\|_v$.
 \end{lemma}
 % \begin{proof}
 %      $ \mc{A}_2 \coloneqq \operatorname{Id}- \mc{R}^\top \Lambda\bs{F}$ is clearly Lipschitz, because $\bs{F}$ is; therefore, for almost all $\x$
 %      \begin{align*}
 %          \| D \mc{A}_2 (x) \|_m &= \| I - \mc{R}^\top\Lambda D\bs{F}(\x) \|_m
 %          \\
 %          & = \max \left \{1, \max_{\i} \| I  - \alpha_i DF(\x)_{i,i} \| + \textstyle \sum_{j\neq i} \| DF(\x)_{i,j} \| \right\} 
 %          \\
 %          & \leq \max \left\{ 1, m
 %          \right\}
 %      \end{align*}
 % \end{proof}
 \begin{proof}
 For any $\x_{i} , \y_{i}\in\R^{n}$, we have 
     \begin{align*}
        & 
        \|x_i-y_i - \alpha_i (\nabla_{x_i}J_i(x_i,\x_{i,-i}) - 
         \nabla_{x_i}J_i(y_i,\y_{i,-i}) )\|
         \\
         {\leq}
         &\|x_i-y_i - \alpha_i( \nabla_{x_i}J_i(x_i,\x_{i,-i}) 
         -\nabla_{x_i}J_i(y_i,\x_{i,-i}))\|\\
         &+
         \alpha_i\| \nabla_{x_i}J_i(y_i,\x_{i,-i}) - 
         \nabla_{x_i}J_i(y_i,\y_{i,-i})\|
         \\
         \overset{(a)}{\leq} &\left(\min\{|1-\alpha_i \mu_i|,|1-\alpha_i \ell_{i,i}| \}\right) \|x_i-y_i \| \\
         &+ \alpha_i  \textstyle \sum_{i,j\in\mc{I},j\neq i} \ell_{i,j} \| \x_{i,j} - \y _{i,j}\|
         \\
         =  &(1-\alpha_i \mu_i)\|x_i-y_i \|  + \alpha_i \textstyle \sum_{i,j\in\mc{I}, j\neq i} \ell_{i,j} \| \x_{i,j} - \y _{i,j}\|
         \\
         \overset{(b)}{\leq} & (1-\alpha_i \mu_i   + \alpha_i \textstyle  \sum_{i,j\in\mc{I}, j\neq i} \ell_{i,j} ) \max_{j\in\mc{I}}{\|\x_{i,j} -\y_{i,j} \|}
         \\
         \overset{(c)}{\leq} &\max_{j\in\mc{I}}{\|\x_{i,j} -\y_{i,j} \|}
     \end{align*}
     where $(a)$ holds because $\min\{|1-\alpha_i \mu_i|,|1-\alpha_i \ell_{i,i}| \}=1-\alpha_i \mu_i$ corresponds to the usual contraction factor for the gradient method \cite[Lemma 10]{Qu2017} (with $\alpha_i <\frac{1}{\ell_{i,i}}$), 
     and by  Lipschitz continuity of $\nabla_{x_i} J_i$; $(b)$ follows because $(1-\alpha_i \mu_i) >0$; 
     and $(c)$ is due to the fact that $1-\alpha_i \mu_i + \alpha_i\sum_{j\neq i} \ell_{i,j} \leq 1$ as specified in \cref{asm:DD_FG}.

     As a consequence, for any $\x,\y\in\R^{Nn}$ defined as in \eqref{eqn:ximat} and \eqref{eqn:xmat}, we can write 
     \begin{align*}
         &
         \| \x-\mc{R}^\top \Lambda \bs{F}\x -\y-\mc{R}^\top \Lambda \bs{F}\y \|_v 
         \\ =& \max_{\i} \bigg( \max  \bigg\{\max_{j \neq i} \| \x_{i,j}-\y_{i,j}\|, \\
         & \|x_i -y_i - \alpha_i\nabla_{x_i} J_i(x_i,\x_{-i})+ \alpha_i\nabla_{x_i} J_i(y_i,\y_{-i}) \|    \bigg \} \bigg)
         \\
         \leq &\max_{\i} \max_{j\in\mc{I}} \|\x_{i,j}-\y_{i,j} \| = \|\x-\y\|_v.
     \end{align*} 
     which completes the proof of \cref{lem:OperatorIdRF}. \hfill 
 \end{proof}
 
As $\mc{A}$ is the composition of three operators that are nonexpansive in $\|\cdot\|_v$, it is also nonexpansive in $\|\cdot\|_v$, which proves the theorem.
\hfill
\end{proof}

\begin{remark}\label{rem-stepsize}
    Besides being much larger than the one that can be provided under the strong monotonicity assumption (without diagonal dominance),  the bound on the step-sizes in \cref{theo:conv} is particularly desirable for two reasons:
    \begin{itemize}
        \item Each agent $i$ can compute the bound on $\alpha_i$ locally.
        \item Remarkably, the bound on each $\alpha_i$ is independent of the communication graph (e.g., the connectivity structure). 
    \end{itemize}
\end{remark}

\mattiasay{
\begin{remark}
    While \cref{asm:Lipschitz} postulates smoothness of the cost functions over the whole space $\R^n$, we can relax this assumption to hold only over $\Omega$ if $\x^0 \in \Omega^N$ (in this case, $\x^k \in \Omega^N$ for all $k$ by the update in \eqref{eq:algo:CASE1}); this can also result in better bounds for the quantities $\mu_i$ and $\ell_{i,j}$. 
\end{remark}
}

\section{Numerical results} \label{sec:simulation}

We evaluate the performance of the proposed approach for the power control problem in optical networks. The details of the problem are described in \cite{Pan2007} and references therein. Specifically, the optical signal-to-noise ratio (OSNR) optimization problem in optical networks is formulated as an $N$-player noncooperative game, based on a general network OSNR model with linear pricing and OSNR-like utility. The cost function $J_i:\Omega_i \to \mathbb{R}$ of each player $i$ is given as follows:
\begin{align}
    \!\!J_i(x) = \eta_ix_i-\b_i\!\left[\ln\left( \!1+a_i\frac{\g_i(x)}{1-\Phi_{ii}\g_i(x)} \!\right)\!-x_i\right]\!,\!\!
\end{align}
where $x_i\in\Omega_i=[x_{\min},x_{\max}]$ denotes the $i$th channel signal power. Here, $\eta_i$, $\b_i$ and $a_i$ represent channel-specific parameters, with $\eta_i$ and $\beta_i$ capturing the price-utility trade-off, while $\gamma_i(x)$ represents the OSNR, defined as
\begin{align}
    \g_i(x)=\frac{x_i}{n_0+\sum_j\Phi_{ij}x_j},
\end{align}
for a given system matrix $\Phi$ and input noise $n_0$. 

The elements of the pseudo-gradient mapping $F(x)$ are:
\begin{align}
    \nabla_{\!x_i}J_i(x)=\eta_i+\b_i-\frac{a_i\b_i}{n_0+\sum_j\Tilde{\Phi}_{ij}x_j},
\end{align}
where $\Tilde{\Phi}_{ij}\!=\!\begin{cases}
        a_i,&\!\!\!i=j\\
        \Phi_{ij},&\!\!\!i\neq j
    \end{cases}$. 
Additionally, the Jacobian matrix $\Theta$ of the pseudo-gradient mapping $F(x)$ is given by
\begin{align}
	\label{eq:Jacobian-pseudo-gradient}
    % &\Theta _{ii} = \nabla^2_{\!x_i}J_i=\frac{a_i^2\b_i}{(n_0+\sum_j\Tilde{\Phi}_{ij}x_j)^2},\quad \quad~       i\in\mc{I},\\
	&\Theta _{ij} = \nabla_{\!x_j}\!\!\nabla_{\!x_i}J_i=\frac{a_i\b_i\Tilde{\Phi}_{ij}}{(n_0+\sum_j\Tilde{\Phi}_{ij}x_j)^2},\quad i,j\in\mc{I}.
\end{align}
If $a_i$ are chosen such that $\Tilde{\Phi}$ is strictly diagonal dominant, i.e.,
\begin{align}\label{eq-Ex-condition}
    a_i > \sum_{j \neq i}\Phi_{ij},
\end{align}
then it can immediately be validated that the Jacobian matrix $\Theta$ is also strictly diagonal dominant.

To demonstrate convergence, we compare the algorithm's results with the actual unique \gls{NE} $x^*$. Typically $\eta_i$ and $\beta_i$ are selected such that an \gls{NE} solution is inner; in this case, the \gls{NE} can is determined by 
% solving $\nabla_{x_i}J_i = 0$ for all $i\in\mc{I}$. The explicit expression for $x^*$ is given by 
\[x^*=\Tilde{\Phi}^{-1}C(\eta), 
\text{where} C(\eta)=\col\left(\! \left(\tfrac{a_i\b_i}{\eta_i+\b_i}-n_0\right)_{\!i\in\mc{I}}\right).\] 

In our simulations, the communication graph is a strongly connected directed graph with self-loops. To ensure strong connectivity, we establish a directed cycle linking all players. We define the row-stochastic weight matrix $W$ as follows
\[w_{ij}=\begin{cases}
0, &\text{if } j\not\in\cNini,\\
\delta, &\text{if } j\in\cNini \text{ and } i\ne j,\\
1-\delta d(i), &\text{if } i=j,
\end{cases}\]
where $d(i)=|\cNini|$  and $\delta=\frac{0.5}{\max_{i}\{d(i)\}}$. The algorithm terminates when $\|\x^{k+1}-\x^k\|<10^{-7}$ or when the iteration limit of $10^7$ is reached. 
% , i.e., $\a_i = \a$ for all $i \in \mc{I}$.  

% and follow the setup in \cite{Pan2007}.
% we consider an optical fiber link with ten amplifiers, each characterized by a parabolic gain shape given by:
% \[G=-4\times 10^{16}\times(\lambda-1555\times 10^{-9})^2+15 \text{dB,}\]
% where $\lambda$ represents the channel wavelength, and there is a span loss of 10 dB.

Consider the six-player case, the parameter values employed are $\eta_i=1$ for all $i \in \mc{I}$, $\beta=[0.5, 0.51, 0.52, 0.3, 0.31, 0.32]$, $a=[0.261, 0.494, 0.107, 0.366, 0.208, 0.305]$, $x_{\min}=0.2$ miliwatt (mW), $x_{\max}=2$ mW and $n_0=0.43\times 10^{-6}$ mW. The system matrix $\Phi$ is given by:
\[\Phi = \begin{bsmallmatrix}
7.463 &7.378 &7.293 &7.210 &7.127 &6.965 \\
7.451 &7.365 &7.281 &7.198 &7.115 &6.953 \\
7.438 &7.353 &7.269 &7.186 &7.103 &6.942 \\
7.427 &7.342 &7.258 &7.175 &7.093 &6.931 \\
7.409 &7.324 &7.240 &7.157 &7.075 &6.914 \\
7.387 &7.303 &7.219 &7.136 &7.055 &6.894
\end{bsmallmatrix}\times 10^{-5}.\]

\begin{figure}[t!]
	\centering
	\subfigure[Algo. 1 ($\gamma=0.2$)]{\includegraphics[width=0.235\textwidth]{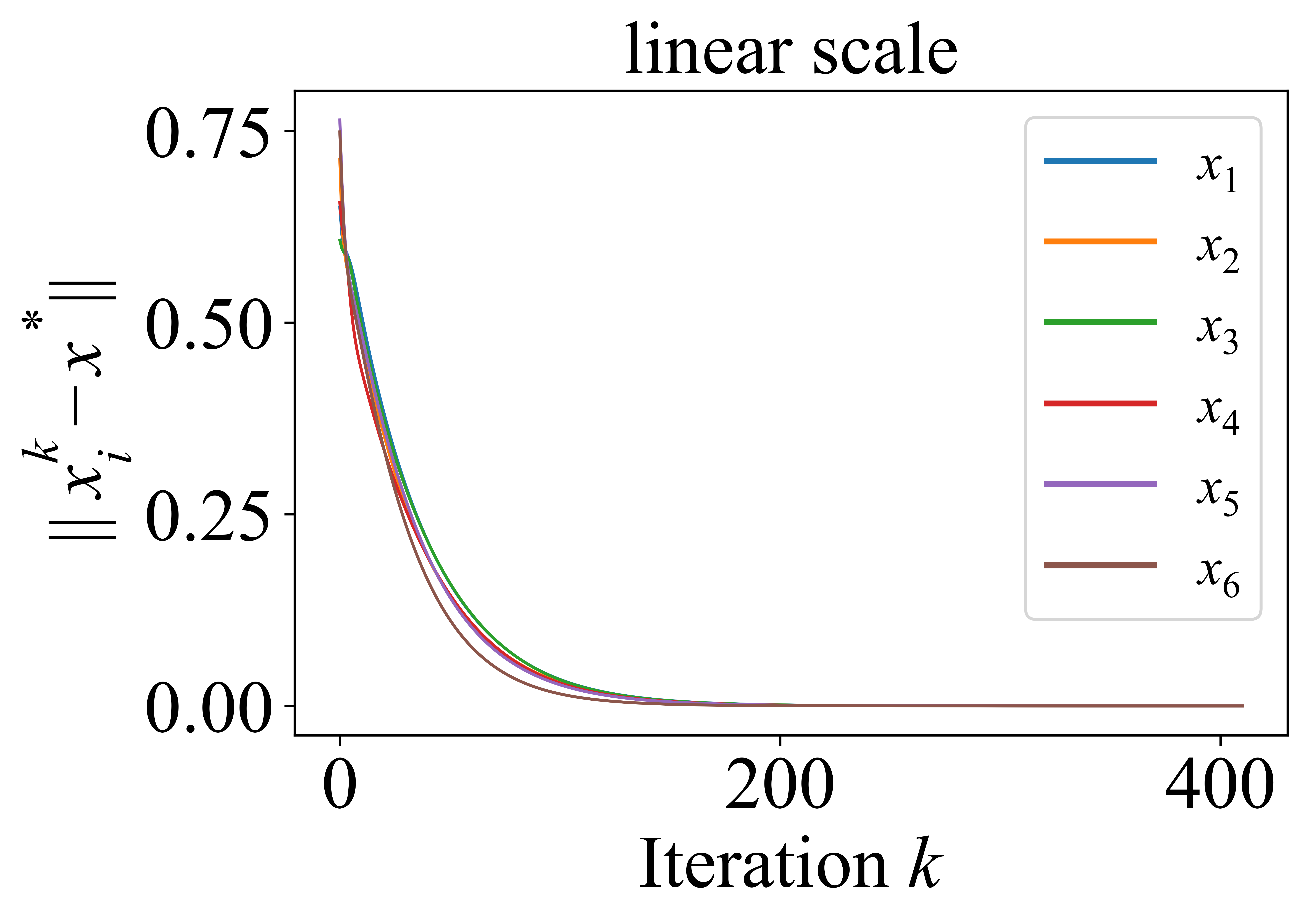}
    \label{fig:ErrorsAgents}}
    \hspace*{-1em} 
    \subfigure[Errors comparison]{\includegraphics[width=0.235\textwidth]{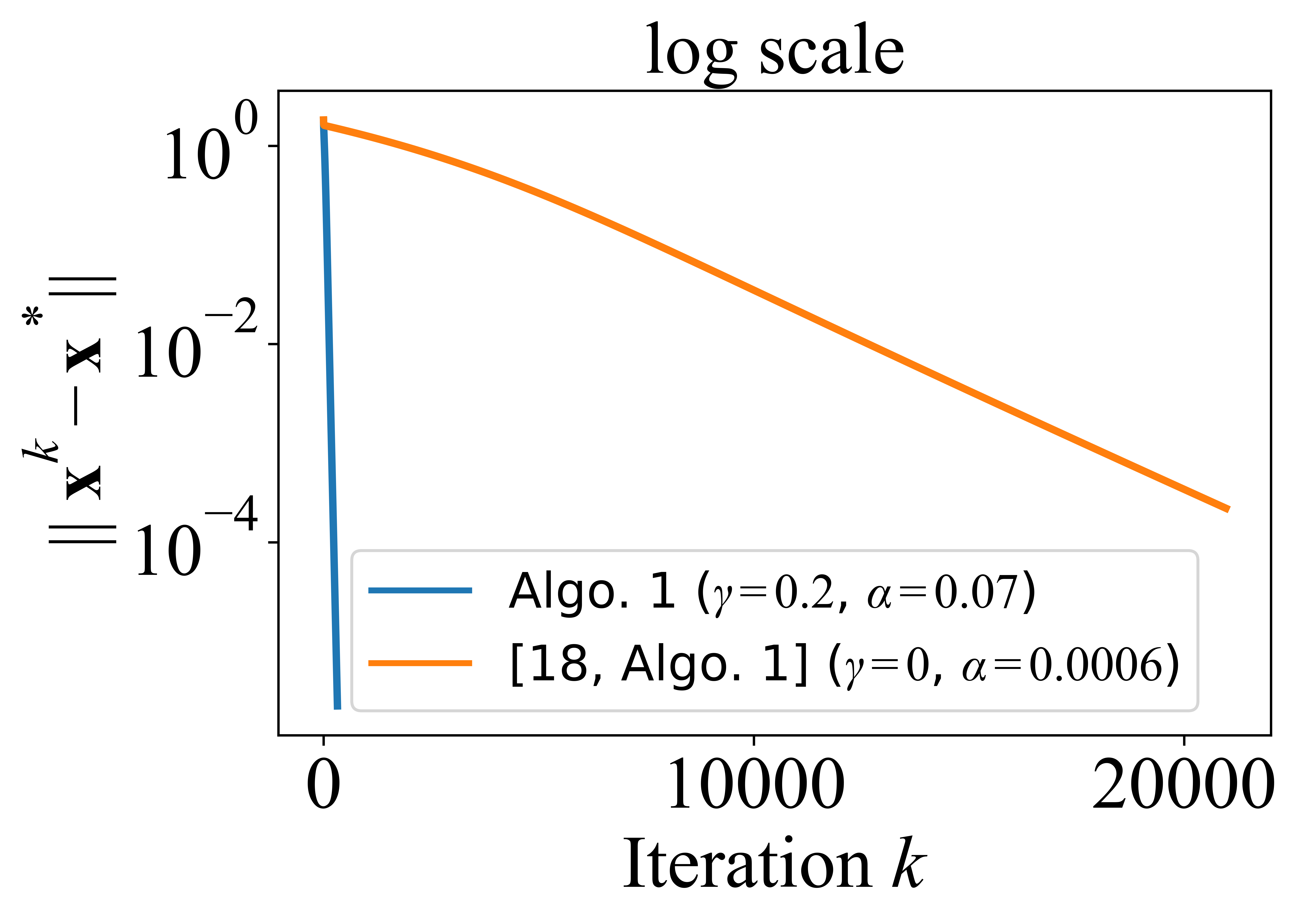}
    \label{fig:ErrorsComparison}}
    \vspace{-0.25cm}
	\caption{Plots of errors.}
	\label{fig:errors}
	\vspace{-0.7cm}
\end{figure}

These parameters satisfy the condition specified in \eqref{eq-Ex-condition} and the $\gls{NE}$ is $x^*=$[$0.3329$, $0.3375$, $0.3412$, $0.2305$, $0.2361$, $0.2421$] mW. The bounds for the step-sizes can be determined using \cref{theo:conv} and \eqref{eq:Jacobian-pseudo-gradient}, resulting in the step-size values $[0.08, 0.07, 0.07, 0.13, 0.12,0.12]$. \cref{fig:ErrorsAgents} shows the convergence result for this game instance.

\begin{table*}[ht]\centering
\vspace{0.21cm}
\begin{tabular}{|c|ccc|ccc|ccc|}
\hline
\multirow{2}{*}{} & \multicolumn{3}{c|}{$\gamma = 0.2$}                                          & \multicolumn{3}{c|}{$\gamma = 0.5$}                                          & \multicolumn{3}{c|}{$\gamma = 0.8$}                                           \\ \cline{2-10} 
& \multicolumn{1}{c|}{Error}      & \multicolumn{1}{c|}{\# Iters}    & \!\!\! Run Time (s) \!\!\! & \multicolumn{1}{c|}{Error}      & \multicolumn{1}{c|}{\# Iters}    & \!\!\! Run Time (s) \!\!\! & \multicolumn{1}{c|}{Error}      & \multicolumn{1}{c|}{\# Iters}     & \!\!\! Run Time (s) \!\!\! \\ \hline
$\!\!N=10\!\!$            & \multicolumn{1}{c|}{$\!\!1.384\!\!\times \!\!10^{-5}\!\!$} & \multicolumn{1}{c|}{$\!\!5454.3\!\!$} & $\!\!0.748\!\!$  & \multicolumn{1}{c|}{$\!\!2.224\!\!\times \!\!10^{-5}\!\!$} & \multicolumn{1}{c|}{$\!\!8623.5\!\!$} & $\!\!1.193\!\!$  & \multicolumn{1}{c|}{$\!\!5.586\!\!\times \!\!10^{-5}\!\!$} & \multicolumn{1}{c|}{$\!\!21049.6\!\!$} & $\!\!2.907\!\!$  \\ \hline
$\!\!N=20\!\!$            & \multicolumn{1}{c|}{$\!\!3.140\!\!\times \!\!10^{-5}\!\!$} & \multicolumn{1}{c|}{$\!\!12925.8\!\!$} & $\!\!4.946\!\!$  & \multicolumn{1}{c|}{$\!\!5.036\!\!\times \!\!10^{-4}\!\!$} & \multicolumn{1}{c|}{$\!\!20445.5\!\!$} & $\!\!7.901\!\!$  & \multicolumn{1}{c|}{$\!\!1.263\!\!\times \!\!10^{-4}\!\!\!\!$} & \multicolumn{1}{c|}{$\!\!49957.6\!\!$} & $\!\!16.664\!\!$  \\ \hline
\end{tabular}
\vspace{-0.1cm}
\caption{Average convergence performance over $1000$ simulations.} \label{table:Simulations} 
\vspace{-0.6cm}
\end{table*}

The step-size bounds in our \cref{theo:conv} are significantly larger than those based on the strong monotonicity assumption. For a practical comparison, we consider the algorithm proposed in \cite{nguyen2022distributed}. Take, for example, a directed cycle graph with self-loops and uniform step-sizes $\a$ for simplicity. In this scenario, the theoretical bound from \cite{nguyen2022distributed} results in a step-size of $\a=0.0006$, which is notably smaller than $\a=0.07$ obtained from \cref{theo:conv}. This conservative estimation is also emphasized in \cite{nguyen2022distributed}. Consequently, 
% although \cite[Algo. 1]{nguyen2022distributed} typically leads to faster convergence, as it corresponds to the case $\gamma=0$ (see \cref{rem-AveragingParam}),
adhering to the theoretical bound in \cite{nguyen2022distributed} results in significantly slower convergence. \cref{fig:ErrorsComparison} visually illustrates our algorithm's faster convergence, primarily attributable to the larger theoretical bound.

Table~\ref{table:Simulations} further compares the average performance of the
proposed algorithm over $1000$ simulations, considering various system parameters and communication networks. As expected, $\gamma$ closer to $0$ improves convergence speed, as players prioritize incoming information over their current estimates.

\section{Conclusions and Future Work}
\label{sec:conc}
In this paper, we have introduced a distributed gradient-based algorithm for seeking \gls{NE} within a directed communication network. This algorithm incorporates an averaging parameter to control the update speed. Convergence is guaranteed under the diagonal dominance assumption of the Jacobian matrix of the pseudo-gradient mapping, with explicit bounds for non-identical step-sizes that remain independent of the network structure. A key avenue for future research involves exploring the convergence rate of the algorithm and examining the impact of \mattiasay{time-varying communication networks}.
% THe original KM allows for time-varying parameters, so we can immediately have that

\bibliographystyle{IEEEtran}
\bibliography{references}

\end{document}

%% file: slide_macros.tex
\usepackage{amsmath,mathtools, amssymb, bbm, xspace}
\usepackage{booktabs,ragged2e}
\usepackage{multirow}
\usepackage{textgreek}
\let\labelindent\relax
\usepackage{enumitem}
\usepackage[capitalize]{cleveref}
\usepackage[flushleft]{threeparttable}
\usepackage{graphicx}
\usepackage[acronym]{glossaries}

\makeglossaries
\newacronym{KKT}{KKT}{Karush--Kuhn--Tucker}
\newacronym{ADMM}{ADMM}{alternating direction method of multipliers}
\newacronym{OPF}{OPF}{optimal power flow}
\newacronym{OPFP}{OPFP}{optimal power flow problem}
\newacronym{NUM}{NUM}{network utility maximization}
\newacronym{LMI}{LMI}{linear matrix inequality}
\newacronym{BMI}{BMI}{bilinear matrix inequality}
\newacronym{LM}{LM}{Lyapunov-Metzler}
\newacronym{SDP}{SDP}{semidefinite programming}
\newacronym{LTI}{LTI}{linear time invariant}
\newacronym{MJLS}{MJLS}{Markov jump linear system}
\newacronym{PID}{PID}{proportional-integral-derivative}
\newacronym{PPA}{PPA}{proximal-point algorithm}
\newacronym{PPPA}{PPPA}{preconditioned proximal-point algorithm}
\newacronym{PPP}{PPP}{preconditioned proximal-point}
\newacronym{NE}{NE}{Nash equilibrium}
\newacronym{GNE}{GNE}{generalized Nash equilibrium}
\newacronym{v-GNE}{v-GNE}{variational GNE}
\newacronym{ISS}{ISS}{input-to-state stability}

\def\diag{\mathop{\hbox{\rm diag}}}

\def\spose#1{\hbox to 0pt{#1\hss}}

\def\text #1{\hbox{\quad#1\quad}}

\def\nthinsp{\mskip -2   mu}

\def\G{_{\scriptscriptstyle G}}

\def\k{_k}

\def\R{_{\scriptscriptstyle R}}

\def\superstar{^{\raise 0.5pt\hbox{$\nthinsp *$}}}
\def\SUPERSTAR{^{\raise 0.5pt\hbox{$*$}}}
\def\lamstarT {\lambda^{\raise 0.5pt\hbox{$\nthinsp *$}T}}

\def\hbar{\skew{4.2}\bar h}

		\def\bkE{{\rm I\kern-.17em E}}
		\def\bk1{{\rm 1\kern-.17em l}}
		\def\bkD{{\rm I\kern-.17em D}}
		\def\bkR{{\rm I\kern-.17em R}}
		\def\bkP{{\rm I\kern-.17em P}}
		\def\bkY{{\bf \kern-.17em Y}}
		\def\bkZ{{\bf \kern-.17em Z}}

		\def\bc{\begin{center}}
		\def\be{\begin{enumerate}}
		\def\bi{\begin{itemize}}
		\def\ec{\end{center}}
		\def\ee{\end{enumerate}}
		\def\ei{\end{itemize}}
		\def\es{\end{small}}
		\def\eS{\end{slide}}

	\def\cp2problem#1#2#3#4{\fbox
		 {\begin{tabular*}{0.9\textwidth}
			{@{}l@{\extracolsep{\fill}}l@{\extracolsep{6pt}}l@{\extracolsep{\fill}}c@{}}
				#1 & & $#4 $ 
			\end{tabular*}}}

		\def\bkE{{\rm I\kern-.17em E}}
		\def\bk1{{\rm 1\kern-.17em l}}
		\def\bkD{{\rm I\kern-.17em D}}
		\def\bkR{{\rm I\kern-.17em R}}
		\def\bkP{{\rm I\kern-.17em P}}
		
		\def\bkZ{{\bf{Z}}}

\newcommand {\beeq}[1]{\begin{equation}\label{#1}}
\newcommand {\eeeq}{\end{equation}}
\newcommand {\bea}{\begin{eqnarray}}
\newcommand {\eea}{\end{eqnarray}}

\def\texitem#1{\par\smallskip\noindent\hangindent 25pt
               \hbox to 25pt {\hss #1 ~}\ignorespaces}

\newcommand{\bs}{\boldsymbol}
\newcommand{\mc}{\mathcal}
\newcommand{\col}{\mathrm{col}}

\newcommand{\beq}{\begin{equation}}
\newcommand{\eeq}{\end{equation}}
\newcommand{\beqn}{\begin{eqnarray}}
\newcommand{\eeqn}{\end{eqnarray}}
\newcommand{\beqno}{\begin{eqnarray*}}
\newcommand{\eeqno}{\end{eqnarray*}}
\newcommand{\bma}{\begin{displaymath}}
\newcommand{\ema}{\end{displaymath}}
\newcommand{\bnu}{\begin{enumerate}}
\newcommand{\enu}{\end{enumerate}}
\newcommand{\bce}{\begin{center}}
\newcommand{\ece}{\end{center}}
\newcommand{\btb}{\begin{tabular}}
\newcommand{\etb}{\end{tabular}}

\def\x{\bs{x}}
\def\y{\bs{y}}

\def\k{{k \in \N}}

\def\i{{i\in\mc{I}}}

\newcommand{\proj}{\operatorname{proj}}
\def\Fbs{\bs{F}}
\def\Rmc{\mc{R}}

\def\R{{\mathbb{R}}}
\def\G{{\mathbb{G}}}

\def\b1{{\mathbf{1}}}

\def\cNini{{\mathcal{N}^{\rm in}_{i}}}
\def\cNouti{{\mathcal{N}^{\rm out}_{i}}}
\def\diag{{\rm diag}}

\usepackage[normalem]{ulem} % to use \sout
\newtheorem{theorem}{Theorem}
\newtheorem{definition}{Definition}

\newtheorem{lemma}{Lemma}
\newtheorem{remark}{Remark}
\newtheorem{standing}{Blanket Assumption}

\newtheorem{assumption}{Assumption}

\usepackage{empheq}
\renewcommand{\emph}{\textit}

\newcommand{\g}{{\mathbf{g}}}

\newcommand{\I}{{\mathcal{I}}}

\def\1{{\bs 1}}

\newcommand{\0}{\bs 0}

\newcommand{\N}{{\mathcal{N}}}

\definecolor{myBlue}{rgb}{0.80,0.85,1.00}
\definecolor{myYellow}{rgb}{0.951,1.000,0.547}

\def\bit{\begin{itemize}}
\def\eit{\end{itemize}}
\def\BEAS{\begin{eqnarray*}}
\def\EEAS{\end{eqnarray*}}

\def\re{{\mathbb R}}

\def\a{\alpha}
\def\b{\beta}

\def\g{\gamma}